\documentclass[aps,prl,twocolumn,groupedaddress,showpacs]{revtex4}

\usepackage{graphicx}

\begin{document}

\title{Multiple Quantum Phase Transitions of Plutonium compounds}
\author{Munehisa Matsumoto$^{1}$, Quan Yin$^{1,2}$, Junya Otsuki$^{3}$,
Sergey Yu. Savrasov$^{1}$}
\affiliation{$^1$Department of Physics, University of California, Davis, California
95616, USA\\
$^2$Department of Physics \& Astronomy, Rutgers University, Piscataway, New
Jersey 08854, USA,\\
$^3$Department of Physics, Tohoku University, Sendai 980-8578, Japan}
\date{\today}

\begin{abstract}
We show by quantum Monte Carlo simulations of realistic Kondo lattice models
derived from electronic--structure calculations that multiple quantum
critical points can be realized in Plutonium--based materials. We place
representative systems including PuCoGa$_{5}$ on a realistic Doniach phase
diagram and identify the regions where the magnetically mediated
superconductivity could occur. Solution of an inverse problem to restore
the quasiparticle renormalization factor for $f$-electrons is shown to
be sufficiently good to predict the trends among
Sommerfeld coefficients and magnetism.
Suggestion on the possible experimental verification for this scenario is given for PuAs.
\end{abstract}

\maketitle







\paragraph{Motivation}

Discovery of unconventional superconductivity
in PuCoGa$_{5}$~\cite{sarrao_2002, nick_2005} opened a new arena for the studies of
strongly-correlated materials. It has the highest superconducting transition
temperature $T_{\mathrm{c}}=18.5$~[K] among $f$-electron-based materials and
it has been discussed to reside somewhere in between the Cerium-based heavy
fermion (HF) superconductors and high-$T_{\mathrm{c}}$ cuprates, where the
latter still challenges theoretical control from first-principles.

In the present work we make predictions
on the magnetism and HF behavior of several Pu compounds
including Pu-115's where the mechanism of possible
magnetically-mediated superconductivity is discussed to be more complicated
than their Cerium counterparts~\cite{flint_2008}. Furthermore, with
experimental challenges such as the self-heating of samples due to the
radiative nature of Pu nuclei, a computational guide should be of help regarding the
determination of the linear coefficient of electronic heat capacity,
so-called Sommerfeld coefficient $\gamma$. Our computational method is
based on a recently-developed 
scheme for realistic Kondo lattice simulations~\cite{mm_2009,mm_2010}
which enabled us to predict the location of magnetic quantum critical point
(QCP)~\cite{sachdev_1999} from electronic
structure calculations~\cite{gabi_2006} for HF materials. 

\begin{figure}[tbp]
\scalebox{0.7}{\includegraphics{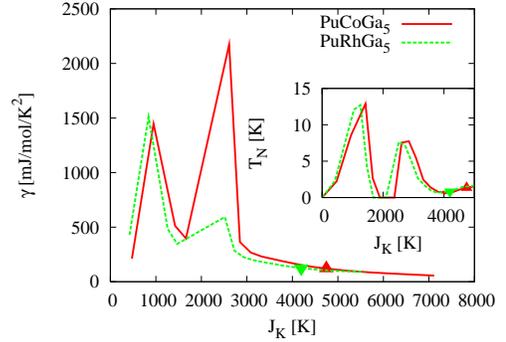}}
\caption{(Color online) Summary of our results for Pu-115's on realistic
Doniach phase diagram. The realistic data point is indicated by the symbol
on the line for each target material.  The main panel show the trend among
Sommerfeld coefficients and the inset show the N\'{e}el temperatures.}
\label{doniach_115s}
\end{figure}
Our main results are shown in Fig.~\ref{doniach_115s} in the format of
Doniach phase diagram~\cite{doniach_1977} plotted with realistic settings
for the target materials. Striking double-dome structure is seen both in the magnetic
Doniach phase diagram of PuCoGa$_{5}$ plotted on $(J_{\mathrm{K}},T_{\mathrm{N}})$-plane
and an analogous plot on $(J_{\mathrm{K}},\gamma )$-plane, where 
$J_{\mathrm{K}}$ is the Kondo coupling and $T_{\mathrm{N}}$ is the Neel
temperature. For Pu-115's there are at least two antiferromagnetic phases
with at least three QCP's. We see that Pu-115's are indeed separated from
the first i.e. lowest-energy magnetic QCP, being consistent with the
situation discussed in Ref.~\cite{flint_2008}. However we find that the
second and third QCP's are encountered on $J_{\mathrm{K}}$-axis and the realistic point
for Pu-115 is actually in the proximity to the latter.

\paragraph{Methods}

The electronic structure calculation based on local density approximation
(LDA) combined with dynamical mean-field theory (DMFT)
(LDA+DMFT) has been successful in addressing interesting properties of
strongly-correlated materials~\cite{gabi_2006}. What motivates us for the
Kondo-lattice model (KLM) description of HF materials is that efficient and
exact quantum Monte Carlo (QMC) simulations in low-temperature region are
possible~\cite{otsuki_2007},
typically around $O(10)$~[K] and down to $O(1)$~[K].
This advantage is due to having only $f$-spins and eliminating
the $f$-electron charge degrees of freedom via Schrieffer-Wolff
transformation~\cite{schrieffer_1966} implemented
in a realistic way~\cite{mm_2009}.
This is in contrast to the fact that the standard LDA+DMFT based
on solving the Anderson model, which
was used e.g. in Ref.~\cite{marianetti_2008} for $\delta $-Pu, typically can reach the temperatures
down to $O(100)$~[K] if the core impurity problem is to be exactly solved by
QMC method. Here, some basis-cutoff schemes have been
implemented~\cite{haule_2007} to reduce the computational cost.

One of the reasons Plutonium compounds have been interesting and difficult
to address is that they reside on the border of itinerancy and localization
of $5f$-electrons among actinides~\cite{moore_2009,santini_1999}.
At least for Pu-chalcogenides and pnictides, experimental evidence
for localized $5f$-electrons was revealed~\cite{gouder_2000} and
a recent theoretical work~\cite{chuck_2010} agrees with that so these can be benchmark cases for the
realistic KLM simulations. For Pu-115's it has been known that the Curie law
persists down to the superconducting temperature
for PuCoGa$_5$~\cite{sarrao_2002} which supports the presence of localized $5f$-electrons, but
some attention must be reserved for a possible sample dependence: radiative
Pu decays into U, which can introduce magnetic impurities.
It is thus controversial whether the Curie-Weiss law is intrinsic
or not~\cite{hiess_2008}. Analyses of experiments point to
$n_f =5.03$ for $\delta$-Pu~\cite{moore_2009}
which we believe is sufficient for the KLM to work.
The valence deduced from the calculations
shows a much larger spread
(between 4~\cite{eriksson_1999} and 6~\cite{havela_2007})
with the values of 5.2 for $\delta$-Pu~\cite{shim_2007}
and 5.26 for PuCoGa$_5$~\cite{pezzoli_2011}
from most accurate CT-QMC calculations.

Our realistic KLM framework for the above-mentioned Pu compounds goes as
follows. At the first stage
LDA for $s$, $p$, and $d$-conduction electrons and
Hubbard-I approximation~\cite{hubbard_1963} for the self-energy of localized 
$f$-electrons gives us the partial densities of states and hybridization
functions as prescribed by LDA+DMFT framework~\cite{gabi_2006}.
The data are summarized in Table~\ref{results_LDA_Hub1}.
It is clear that Pu-115's have much higher energy scales than the Cerium ones~\cite{mm_2010}.
At the second stage we solve the low-energy effective KLM Hamiltonian with
dynamical-mean field theory~\cite{georges_1996,otsuki_2009_formalism},
utilizing state-of-the-art continuous-time quantum Monte Carlo (CT-QMC) impurity
solver~\cite{werner_2006,haule_2007,otsuki_2007}.
For the $5f$-orbitals of
Pu, it is known that there is a big spin-orbit splitting of $1$~[eV] and the
five possibly localized electrons fill in the lower $j=5/2$ multiplet up to
leaving one localized hole~\cite{cooper_1983}. Photoemission experiments show that the level of
the localized hole is separated from the Fermi level by $1$~[eV] which
is verified by our theoretical estimates.
We neglect the crystal-field splittings which are known to be small in Pu
compounds in the local $5f$ level.
\begin{table}
\begin{tabular}{cccc}
material & $N^{\rm tot}(0)$ & $N^{f}(0)$ & $-{\rm Tr}\Im\Delta(0)/\pi$~[eV] \\ \hline
PuCoGa$_5$ & 34.33\ldots & 1.203\ldots & 0.705  \\
PuRhGa$_5$ & 33.03\ldots & 1.494\ldots & 0.912 \\ \hline
$\delta$-Pu & 20.40\ldots & 2.781\ldots & 1.13 \\
PuSe & 17.30\ldots & 4.342\ldots  & 0.619  \\
PuTe & 13.45\ldots & 0.7729\ldots & 0.360  \\
PuAs & 8.351\ldots & 0.6068\ldots & 0.255   \\
PuSb & 10.01\ldots & 0.2699\ldots & 0.193  \\
PuBi & 7.723\ldots & 0.2534\ldots & 0.133  \\ \hline
\end{tabular}
\caption{\label{results_LDA_Hub1} Summary of
LDA+Hubbard-I results for target materials.
The unit of density of states, $N^{\rm tot}(0)$ for all electrons
and $N^{f}(0)$ for $f$-electrons on the Fermi level, is [states/Ry/cell].}
\end{table}

\paragraph{Solving an inverse problem to restore $f$-electrons}

Even if we eliminated $f$-electrons and kept only $f$-spins
in our KLM, part of the information for the
localized $f$-electrons can be restored from the relation 
\(
\Sigma_{c}(i\omega_n)\equiv
V^2/[i\omega_n-\epsilon_{f}-\Sigma_{f}(i\omega_n)],
\)
where $\Sigma_{c}$ is our conduction-electron
self energy, $i\omega_n=(2n+1)\pi T$ is the Matsubara frequency, $\epsilon_{f}=-1$~[eV] is
the position of the impurity level,
and  $\Sigma_{f}$ is the $f$-electron self-energy which we do not have explicitly
in our KLM calculations.
Provided that
we reach the temperature for a given target material to be in a Fermi-liquid region
concerning its $f$-electrons, which is mostly the case for Pu compounds,
the quasiparticle renormalization factors are well defined and written as
\(
z_{x}=\left(1-\partial\Im\Sigma_{x}(i\omega_n)/\partial (i\omega_n)\right)^{-1},
\)
with $x=c$ and $f$ for conduction electrons and $f$-electrons, respectively.
Here the derivative is taken at $i\omega_{n}=0$. We get from the above
definition of $\Sigma_{c}$ the following inversion relation :
\(
z_f=[|\Sigma_c(0)|^2/V^2]z_{c}/(1-z_{c}).
\)
Because $z$'s are written in terms of the derivative of the corresponding
self-energy at the lowest frequency, our effective low-energy description
based on KLM enables a good solution of this inverse problem as far as $z_f$
is concerned. The Sommerfeld coefficient $\gamma=(1/3)\pi^{2} N_{\mathrm{eff}}(0)$,
where $N_{\mathrm{eff}}(0)$ is the effective total density of states (DOS) on the
Fermi level, can be estimated by
\(
N_{\mathrm{eff}}(0)=N_{c}(0)/z_{c}+N_{f}(0)/z_{f},
\)
where $N_{c}(0)$ is DOS of $s$, $p$, $d$-conduction electrons and $N_{f}(0)$
is that of localized $f$-electrons in our LDA+Hubbard-I calculations. With a
given KLM, we extract $z_{c}$ from $\Sigma_{c}$ obtained after DMFT, invert
it to $z_{f}$, and get the Sommerfeld coefficient $\gamma$ with the above formula.
In this way we can restore an
analogue of Doniach phase diagram for $\gamma$ as was shown
in Fig.~\ref{doniach_115s} for Pu-115's. The results on the realistic data point for
each target material are summarized in Table~\ref{results_KLM} together
with the experimental data taken from the literature. Our prediction follows
the experimental trend among $\gamma$ semi-quantitatively.
We note that $\gamma$ is sensitive
to the estimate of the realistic point
of $J_{\mathrm{K}}$ especially around QCP's,
considering the sharp peak structure as seen in Fig.~\ref{doniach_115s} for
the plot of $\gamma$ vs $J_{\mathrm{K}}$. So the overall trend among
materials is the most important result.
\begin{table}
\begin{tabular}{ccccc}
material & $z_f$ & $z_c$ & our $\gamma$ & experimental $\gamma$ \\ \hline
PuCoGa$_5$  &  0.0403 & 0.0496 & 120 &  80$^{a}$-116$^b$  \\
PuRhGa$_5$  &  0.0367 & 0.0462 & 130 & 50$^a$-80$^c$ \\ \hline
$\delta$-Pu &  0.0625 & 0.0738 & 49 &   50-64$^d$\\
PuSe &  0.0480 & 0.0232 & 110 & 90$^e$ \\
PuTe &  0.00883 & 0.0313 & 85 & 30$^f$-60$^e$ \\
PuAs &  0.0588  & 0.00456  & 295 &  \\
PuSb &  0.0231 & 0.0168  & 102 & 6$^{b}$-20$^g$ \\
PuBi &  0.00202 & 0.0965 & 35 &  \\ \hline
\end{tabular}
\\
$^a$ Ref.~\cite{thompson_2007}
$^b$ Ref.~\cite{javorsky_2005}
$^c$ Ref.~\cite{sarrao_2007}
$^d$ Ref.~\cite{lashley_2003}
$^e$ Ref.~\cite{fournier_1990}
$^f$ Ref.\cite{stewart_1991}
$^g$ Ref.\cite{hall_1986}
\caption{\label{results_KLM} Summary of our data obtained with
realistic Kondo lattice simulations and our prediction for $\gamma$ based on them.
The unit of $\gamma$ is [mJ/mol/K$^{2}$].
Experimentally known results are taken from the literature.}
\end{table}

\paragraph{Magnetism and quasiparticle renormalizations}

\begin{figure}[tbp]
\scalebox{0.4}{\includegraphics{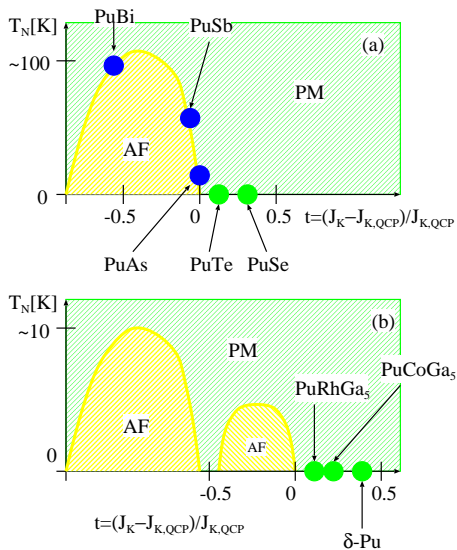}}
\caption{(Color online) Schematic summary
of our magnetic phase diagrams for Pu compounds
plotted on the $(t,T_{\mathrm{N}})$-plane
where $t\equiv (J_{\mathrm{K}}-J_{\mathrm{K,QCP}})/J_{\mathrm{K,QCP}}$ is
the rescaled Kondo coupling with $J_{\mathrm{K,QCP}}$ being the first QCP
in (a) and the third QCP in (b).
}
\label{doniach_pu2}
\end{figure}

The results for magnetism are schematically summarized
in Fig.~\ref{doniach_pu2} for all target materials in the format of a rescaled Doniach
phase diagram. It illustrates how we understand the results in the inset of
Fig.~\ref{doniach_115s} for Pu-115's.
Striking multi-dome structure shows up
together with multiple QCP's for materials with strong Kondo coupling. 
We find that Pu-115's are located in a region where antiferromagnetic
long-range order is suppressed, possibly near a hidden or pseudo-QCP, within some
numerical noise at the lowest reachable temperatures at
present. Inspecting the distribution of materials around the QCP's
in Fig.~\ref{doniach_pu2}, we have pnictides on the left-hand side and
chalcogenides on the right-hand side of the antiferromagnetic QCP. This is
consistent with what has been known experimentally, that is, pnictides such
as PuAs, PuSb~\cite{burlet_1984}, and PuBi~\cite{mattenberger_1986} are
magnets and chalcogenides such as PuSe and PuTe
are paramagnets~\cite{fournier_1990}. The actual magnetism is strongly spatially
anisotropic~\cite{santini_1999,mattenberger_1986} whose treatment
is unfortunately beyond the level of single-site DMFT description.
For now we will leave the issue of ordering wavevectors for future projects
and focus on the trends across target materials spanning between magnetism and HF behavior.
The characteristic energy scales of Kondo-screening
and magnetic ordering have been captured by fully
incorporating the frequency-dependence of the hybridization.

\begin{figure}[tbp]
\begin{tabular}{l}
(a) \\
\scalebox{0.6}{\includegraphics{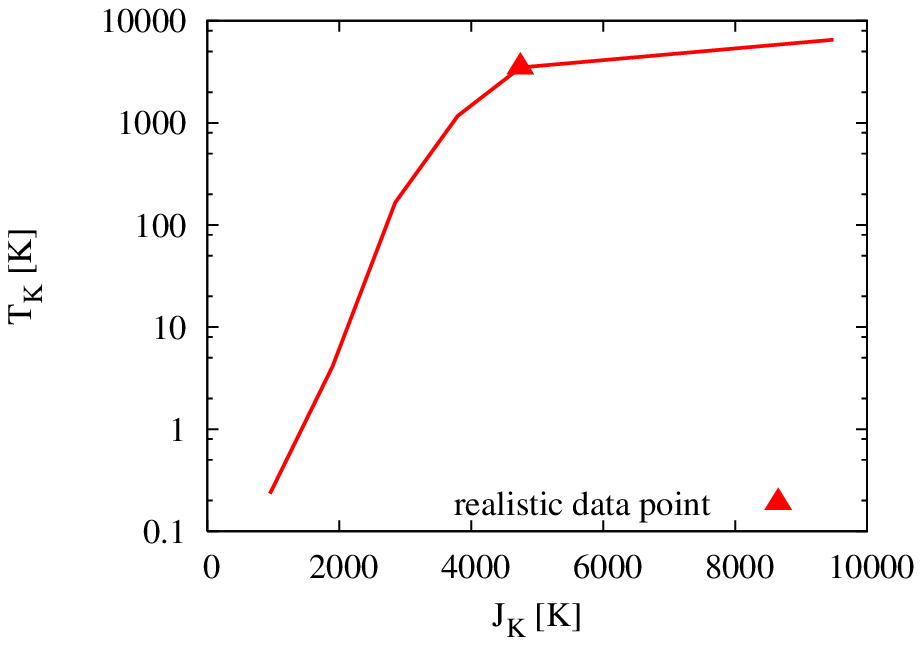}} \\
(b) \\
\scalebox{0.6}{\includegraphics{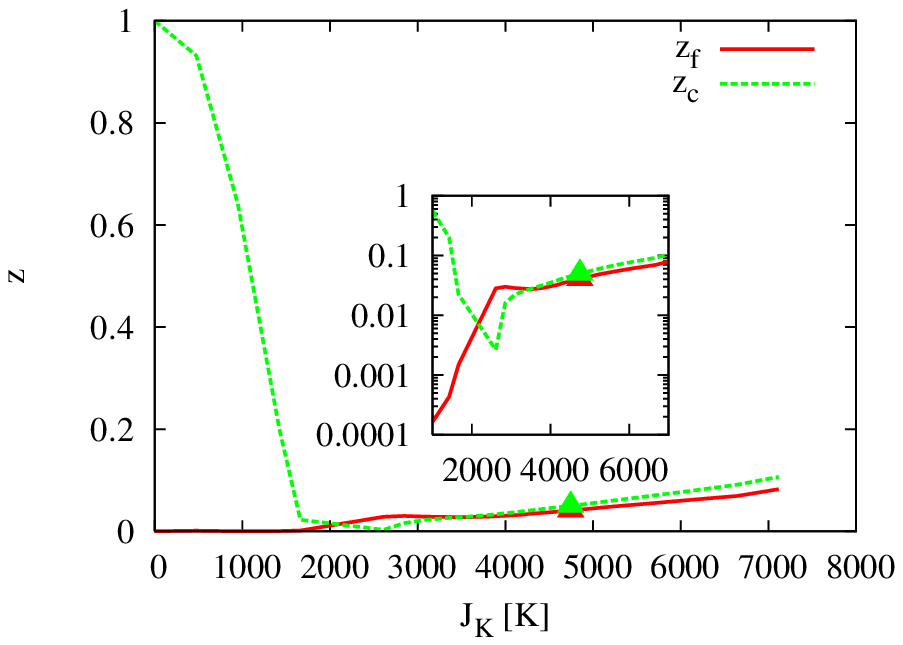}}
\end{tabular}
\caption{(Color online) Analogue of Doniach phase diagram for (a) Kondo temperature
and (b) quasiparticle renormalization factors for PuCoGa$_5$. The inset in (b)
is a zoom-up picture around the first QCP with the vertical axis plotted
in logarithm.}
\label{TKandz}
\end{figure}
Multi-dome structure together with multiple QCP's 
shown in Fig.~\ref{doniach_115s} for Pu-115's can be understood
in terms of strong-coupling nature of the Kondo lattice~\cite{lacroix_1985}
based on the growth of characteristic Kondo energy scale $T_{\mathrm{K}}$
with respect to $J_{\rm K}$
as shown in Fig.~\ref{TKandz}~(a) which is obtained
from our local susceptibility data.
Simple perturbative arguments also prompt for two crossover points
between $T_{\rm RKKY} \sim J_{\rm K}^2\rho$ and $T_{\rm K}\sim \exp[-1/(J_{\rm K}\rho)]/\rho$ temperature scales,
where the latter can saturate at some point with respect to large $J_{\rm K}$ while
the former keeps on growing. Here $\rho$ is the characteristic DOS.

We demonstrate our predictive power regarding
$T_{\mathrm{K}}$ also for the case of $\delta $-Pu where
we get $T_{\mathrm{K}}\sim 10^{3}$~[K] from our local susceptibility data which is
seen to be close to the previous results, $T_{\mathrm{K}}\sim 700$~[K] in
Ref.~\cite{shim_2007}. We note that the Kondo screening energy scale
is approaching a comparative scale to the characteristic bandwidth,
or the kinetic energy of the conduction electrons which is $O(1)$~[eV].
Such situation had been discussed in the literature~\cite{lacroix_1985,sigrist_1992}
in the context of models, which is now found to be realized
in Plutonium heavy-fermion materials.

The behavior of quasiparticle renormalization factors $z_{x}$ ($x=c$ or $f$)
as shown in Fig.~\ref{TKandz}~(b) further
gives the physical picture: starting from $J_{\rm K}=0$
where there are free conduction electrons and completely localized $f$-electrons with
$(z_{c},z_{f})=(1,0)$, the former gradually gets renormalized toward
the first QCP and $f$-electrons gets ``delocalized'' in the sense
that they start to take part in Fermi surface (FS)~\cite{martin_1982,otsuki_2009_large_fs}.
Passing the first QCP, heavy quasiparticles composed both of conduction electrons
and $f$-electrons evolve together after $z_{c}$ has shown a dip around the first QCP,
letting $f$-spins show up again with the underscreening effects that
correspond to the slightly elevated $z_{c}$. Thus after the revival of magnetism
the same thing can happen again and could repeat itself, with the re-defined
much smaller energy scales every time, all the way to the $J_{\rm K}\rightarrow\infty$
limit, being consistent with the phase diagram obtained in Ref.~\cite{lacroix_1985}.


This strong-coupling KLM scenario for Pu compounds can in principle
be checked by de Haas-van Alphen experiments which would
measure the size of the FS
to see if it counts the number of \textquotedblleft localized\textquotedblright\ $f$-electrons.
The ferromagnetic phase of PuAs should be the one with the
\textquotedblleft large\textquotedblright\ FS including the spins of
localized $5f$-electrons. This phase would be in contrast to the typical
magnetic phases in HF compounds with the \textquotedblleft
small\textquotedblright\ FS, being located in the weak-coupling region.
Following the method of Ref.~\cite{otsuki_2009_large_fs}, we can track the
evolution of large FS for representative Pu compounds
obtained from $\left. -\Re \Sigma _{c}(i\omega _{n})\right\vert _{i\omega_{n}=0}$
and we find that PuAs shows a remarkable evolution of ``large'' FS,
which is to be compared with experiments to see
if the strong-coupling KLM picture can hold. 


\paragraph{Conclusions}

We have found that multiple QCP's show up for Plutonium-based materials that
have stronger Kondo couplings than their Cerium counterparts. Our
methodology captures the quasiparticle renormalization factor and the
characteristic energy scale correctly. The striking multi-dome feature of
the Doniach phase diagrams for Pu-115's and $\delta$-Pu
as well as the magnetic and HF behavior among Pu pnictides
and chalcogenides is understood on the basis of strong-coupling limit of KLM.
This picture can be verified by looking at the size of the Fermi surface for
PuAs.

\begin{acknowledgments}
The authors thank E.~D.~Bauer, A.~V.~Chubukov,
P.~Coleman, N.~J.~Curro, R.~Dong,
M.~J.~Han, K.~Haule, K.~Kim, G.~Kotliar,
H.~Shishido, Y.-F.~Yang, C.-H.~Yee for discussions. The present numerical
calculations have been done on \textquotedblleft Chinook\textquotedblright\
in Pacific Northwest National Laboratory. This work was supported
by DOE NEUP Contract No.~00088708.
\end{acknowledgments}

\end{document}